\documentclass [12pt,a4paper]{article}
\usepackage{times}

\DeclareFontFamily{OT1}{times}{}
\DeclareFontShape {OT1}{times}{m }{n }{ <-> ptmr }{}
\DeclareFontShape {OT1}{times}{bx}{n }{ <-> ptmb }{}
\DeclareFontShape {OT1}{times}{m }{it}{ <-> ptmri}{}
\DeclareFontShape {OT1}{times}{bx}{it}{ <-> ptmbi}{}
\usepackage{amsmath}
\usepackage{amsfonts}
\usepackage{amssymb}
\newcommand{\HU}{\text{H}}            
\newcommand{\DUP}{\delta}             
\newcommand{\UPS}{\Upsilon}           
\newcommand{\rmD}{\mathrm{D}}         
\newcommand{\DEF}{:=}                 
\newtheorem{definition}{Definition}   
\newtheorem{theorem}{Theorem}         
\newcommand{\OOO}{\mathrm{O}}         
\newcommand{\ASS}{\asymp}             
\renewcommand{\iiint}{\int\kern -.8em\int\kern -.8em\int}    
\def\Me{{\text{\emph{M}}}}    
\def\Mt{{\text{M}}}           
\def\limea{\lim_{\substack{
                \epsilon \rightarrow 0\\
                       a \rightarrow 0}}}
\def\lime{\lim_{\epsilon \rightarrow 0}}
\def\lima{\lim_{       a \rightarrow 0}}
\numberwithin{equation}{section}
\begin{document}
\title{\bf\vspace{-3.5cm} The classical point-electron in Colombeau's
                          theory of nonlinear generalized functions}

\author{
         {\bf Andre Gsponer}\\
         {\it Independent Scientific Research Institute}\\ 
         {\it Oxford, OX4 4YS, United Kingdom}
       }

\date{J. Math. Phys. {\bf 49} (2008) 102901 \emph{(22 pages)}}

\maketitle

\begin{abstract}

The electric and magnetic fields of a pole-dipole singularity attributed to a point-electron-singularity in the Maxwell field are expressed in a Colombeau algebra of generalized functions.  This enables one to calculate dynamical quantities quadratic in the fields which are otherwise mathematically ill-defined:  The self-energy (i.e., `mass'), the self-angular momentum (i.e., `spin'), the self-momentum (i.e., `hidden momentum'), and the self-force.  While the total self-force and self-momentum are zero, therefore insuring that the electron-singularity is stable, the mass and the spin are diverging integrals of $\delta^2$-functions.  Yet, after renormalization according to standard prescriptions, the expressions for mass and spin are consistent with quantum theory, including the requirement of a gyromagnetic ratio greater than one.  The most striking result, however, is that the electric and magnetic fields differ from the classical monopolar and dipolar fields by $\delta$-function terms which are usually considered as insignificant, while in a Colombeau algebra these terms are precisely the sources of the mechanical mass and spin of the electron-singularity.

\end{abstract}

\section{Introduction} 
\label{int:0}

Colombeau's theory of generalized functions \cite{COLOM1984-,COLOM1985-,COLOM1990-} is now considered by many mathematicians as the preeminent generalization of the theory of distributions such that multiplication is always possible. There is a growing mathematical literature (Ref.~\cite{GROSS2001-, KUNZI2007} and numerous references therein) showing how Colombeau algebras can effectively be applied to the study of singular problems that involve differentiation combined with non-linear operations, such as non-linear partial differential equation, non-smooth differential geometry, generalized Fourier transform, etc. 

    Because of their optimal properties for a wide range of applications, Colombeau algebras have also found numerous applications in classical physics \cite{SCHME1990-,COLOM1992-}, especially in hydrodynamics \cite{COLOM1988-,HU---1998-,BERNA2001-,BATY-2008-} and general relativity \cite{GROSS2001-,STEIN1997-,KUNZI1999-,STEIN2006-,CASTR2008-}.

  There is however comparatively little use of them in electrodynamics \cite{HORMA1998-,HORMA2000-,GSPON2004D,GSPON2007C,GSPON2006B}, despite that products of distributions arise naturally in that domain, especially when attempting to calculate the self-interaction terms arising in any problem dealing with point-like charged particles, e.g., the self-force and the self-energy, which are well-known to be divergent quantities.  In fact, in the few cases in which generalized functions other than Schwartz distributions have been used, the objective has been to provide a better mathematical justification for already known results \cite{ROSEN1981-,ROSEN1982-}.

  In this paper, however, the objective is to reconsider some of the foundations on which calculations with point singularities are made in classical electrodynamics.  Indeed, Maxwell's theory is basically a continuum theory, and the introduction of point charges requires additional assumptions to deal with them.  In this paper, the basic assumption is that the electron is a true pole-dipole singularity of electric charge $e$ and magnetic moment $\vec{\mu}$, and that this singularity is a distribution embedded in a Colombeau algebra.  In fact, that distribution is defined exactly as in distribution theory \cite{TEMPL1953-}, the essential difference being that embedded in a Colombeau algebra it becomes a generalized function (in short, a $\mathcal{G}$-function) which can be manipulated in non-linear expressions as if it were an ordinary $\mathcal{C}^\infty$ function.

  This implies, for instance, that the Coulomb field has a more complicated structure than in the standard formulations of classical electrodynamics, i.e., \cite[p.\,519]{TANGH1962-}, \cite{GSPON2004D},
\begin{equation}
\label{int:1}
  \vec{E}(\vec{r}\,) = \frac{e\vec{r}}{r^3} \UPS (r)
                      - \frac{e\vec{r}}{r^2}   \UPS'(r),
\end{equation}
where $\UPS$ is a $\mathcal{G}$-function (defined in Sec.\,\ref{ups:0}) with properties similar to Heaviside's step function, so that the $\mathcal{G}$-function $\UPS'$ has properties similar to Dirac's $\delta$-function.   Thus, the classical Coulomb field is retrieved if $\UPS=1$, in which case $\UPS'=0$.  In distribution theory, however, the $\UPS$ function is replaced by Heaviside's step function $\HU$, more precisely, by $\lima \HU(r-a)$, whereas what would correspond to $\UPS'$ is simply discarded \cite[p.\,144]{TEMPL1953-}.  For that reason, essential information related to the nature of the Coulomb singularity is lost, which implies that the multiplication of expressions containing the Coulomb field or its derivatives do not make sense mathematically in distribution theory.

   Applying the same principles to the magnetic dipolar field, which is well-known to very precisely characterize the intrinsic magnetic dipole moment of elementary particles, one finds \cite{GSPON2004D}
\begin{equation}
\label{int:2}
    \vec H(\vec{r}\,)
      = \frac{3\vec{r} (\vec\mu\cdot\vec{r}) - \vec\mu}{r^5}\UPS (r)
      + \frac{ \vec{r} \times (\vec\mu\times\vec{r})  }{r^4}\UPS'(r),        \end{equation}
where in contrast to that in \eqref{int:1}, the additional $\delta$-like contribution is well known.  Indeed, its integral is essential in calculating the hyperfine splitting of atomic states \cite[p.\,184]{JACKS1975-}, \cite{JACKS1977-}.

   The pole-dipole singularity assumed to represent the electromagnetic fields of a point electron is therefore completely defined by the fields (\ref{int:1}--\ref{int:2}).  The main goals of the paper are thus to justify them rigorously, and to derive from them the basic dynamical properties of the electron, namely its mass and spin --- as well as to verify that the electron singularity is stable, and that everything is consistent with classical electrodynamics and quantum theory.

   In Secs.\,\ref{col:0} and \ref{imd:0} we give a brief introduction to Colombeau algebras.

 In Sec.\,\ref{poi:0} the general method for introducing point singularities in distribution theory and in Colombeau algebras is presented.  The case of the electric monopole is worked out in details:  Starting from the Coulomb potential, the electric field strength and the charge density are calculated in classical, distributional, and Colombeau theories. 

   In Sec.\,\ref{ups:0} a general calculus, based on a $\mathcal{G}$-function $\UPS$, is developed in order to simplify nonlinear calculations with electromagnetic fields and related quantities arising from point singularities.

   In Secs.\,\ref{mon:0} to \ref{sam:0} the basic dynamical quantities associated to an electron singularity at rest are calculated.  In these calculations the additional $\delta$-terms in (\ref{int:1}--\ref{int:2}) play an essential role, and lead to expressions for the mass and spin which are integrals of $\delta^2$-expressions as in quantum field theory.  In the case of the self-force and of the self-momentum the integrands contain a prevailing $\delta^2$-term but the symmetry of the electric and magnetic fields is such that the angular integrations give zero, so that the electron singularity is stable.

   In Secs.\,\ref{ren:0} and \ref{cqt:0} it is shown that mass and spin are renormalizable according to the usual philosophy of renormalization, and that all calculations and results obtained in the paper are consistent with quantum theory.

   Finally, in Sec.\,\ref{con:0}, we conclude by summarize the main results, to which we could add that these results confirm one of our major motivations, namely that our approach to point-singularities in the Maxwell field may have implications beyond classical physics.  This is because Maxwell's theory, and the concept that truly elementary particles are field-singularities, are integral parts of quantum theory and quantum field theory, which should therefore also be reconsidered in the framework of Colombeau algebras.

\section{Colombeau generalized functions} 
\label{col:0}

A Colombeau algebra $\mathcal{G}$ is an associative differential algebra containing the space $\mathcal{D}'$ of distributions as a subspace.   One calculates with $\mathcal{G}$-functions as with $\mathcal{C}^\infty$-functions.  Colombeau and others have introduced a number of variants of $\mathcal{G}$ but all `Colombeau algebras' have in common one essential feature: The $\mathcal{C}^\infty$ functions are a faithful differential subalgebra of $\mathcal{G}$. 

The elements $g$ of $\mathcal{G}$ are `moderate' one-parameter families $\{g_\epsilon \}$ of $\mathcal{C}^\infty$ functions denoted by $\mathcal{E}_\Mt$, modulo `negligible' families denoted by  $\mathcal{N}$, where the adjectives refer to certain growth conditions in the parameter $\epsilon$ of the family.  More precisely, let $\Omega$ be an open set in $\mathbb{R}^n$, let $\epsilon \in ]0,1[$ be a parameter, and let $\rmD^\alpha$ be any partial differentiation operator where $\alpha$ is a multi-index:
\begin{definition}  The differential algebra of {moderate functions} is
\begin{align}
\nonumber
    \mathcal{E}_\Me(\Omega) \DEF
 \Bigl\{ f_\epsilon : ~ & \forall K \text{~compact in~} \Omega,
                            \forall \alpha \in \mathbb{N}_0^n,\\
\nonumber
         & \exists N \in \mathbb{N}_0 \text{~such that}, \\
\label{col:1}
         & \sup_{\vec{x} \in K} ~ |\rmD^\alpha f_\epsilon(\vec{x}\,)|
         = \OOO(\frac{1}{\epsilon^N})
           \text{~~~as~} \epsilon \rightarrow 0
  \Bigr\},
\end{align}
\end{definition}
\begin{definition} The differential ideal of {negligible functions} is
\begin{align}
\nonumber
    \mathcal{N}(\Omega) \DEF
 \Bigl\{ f_\epsilon : ~ & \forall K \text{~compact in~} \Omega,
                            \forall \alpha \in \mathbb{N}_0^n,\\
\nonumber
         & \forall q \in \mathbb{N}, \\
\label{col:2}
         & \sup_{\vec{x} \in K} ~ |\rmD^\alpha f_\epsilon(\vec{x}\,)|
         = \OOO(\epsilon^q)
           \text{~~~as~} \epsilon \rightarrow 0
  \Bigr\}.
\end{align}
\end{definition}
Then, if we conventionally write $\mathcal{N}$ for any negligible function, we have
\begin{align}
 \label{col:3}
   \forall g_\epsilon, \forall h_\epsilon \in \mathcal{E}_\Mt, \qquad
   ( g_\epsilon + \mathcal{N} ) \cdot
   ( h_\epsilon + \mathcal{N} )
  = g_\epsilon \cdot h_\epsilon + \mathcal{N},
\end{align}
and it is not difficult to prove that the quotient
\begin{align}
 \label{col:4}
        \mathcal{G} \DEF \frac{\mathcal{E}_\Mt}{\mathcal{N}},
\end{align}
is a differential algebra.  That is, an element $g \in \mathcal{G}$ is an equivalence class $[g] = [g_\epsilon + \mathcal{N}]$ of an element $g_\epsilon \in \mathcal{E}_\Mt$, which is called a \emph{representative} of the \emph{generalized function} $g$.  If `$\odot$' denotes multiplication in $\mathcal{G}$, the product $g \odot h$ is defined as the class of $g_\epsilon \cdot h_\epsilon$ where $g_\epsilon$ and $h_\epsilon$ are (arbitrary) representatives of $g$ and $h$; similarly $\rmD g$ is the class of $\rmD g_\epsilon$ if $\rmD$ is any partial differentiation operator.  Therefore, when working in $\mathcal{G}$, all algebraic and differential operations (as well as composition of functions, etc.) are performed component-wise at the level of the representatives $g_\epsilon$.

The specific feature of \emph{Colombeau algebras} which leads a mathematically consistent way of multiplying distributions while insuring that $\mathcal{C}^\infty$ is a faithful differential subalgebra is the use of a special regularization to mollify the distributions.  That is, the distributions $f \in \mathcal{D}'$ are embedded in $\mathcal{G}$ as the convolution\footnote{This definition due to Colombeau differs by a sign from the usual definition of regularization.} 
\begin{align}
\nonumber
          f_\epsilon(x)  \DEF \eta_\epsilon(-x) \ast f(x)
                     &= \int dy~ \frac{1}{\epsilon}
                        \eta\Bigl(\frac{y-x}{\epsilon}\Bigr) ~f(y)\\
\label{col:5}   
                \,\, &= \int dz~\eta(z) ~f(x + \epsilon z),
\end{align}
where the \emph{Colombeau mollifier} $\eta \in \mathcal{S}$ has all moments vanishing, i.e.,
\begin{equation} \label{col:6}
    \Bigl\{
    \int dz~\eta(z) = 1,
    \quad \text{and} \quad
    \int dz~z^n\eta(z) = 0, \quad \forall n=1,...,q \in \mathbb{N}
    \Bigr\}.
\end{equation}
To these minimal conditions additional requirements can be added depending upon the application.  In the present one we require that $\eta(-z)=\eta(z)$, which implies that the embedded Dirac $\delta$-function is even.  Then, the Colombeau embeddings \eqref{col:5} of the Dirac and Heaviside functions are,
\begin{equation} \label{col:7}
     \delta_\epsilon(x) 
      = \frac{1}{\epsilon} \eta\Bigl(-\frac{x}{\epsilon}\Bigr),
    \qquad \text{and} \qquad
 \HU_\epsilon(x) = \int_{-\infty}^{x/\epsilon} dz~\eta(-z).
\end{equation}
which are moderate functions with $N=1$ and $0$, respectively.  More generally:
\begin{theorem}[Colombeau local structure theorem]\label{theo:3}
Any distribution is locally a moderate generalized function \emph{\cite[p.\,61]{COLOM1984-}}, \emph{\cite[p.\,62]{GROSS2001-}}.
\end{theorem} 
Colombeau proved that the set \eqref{col:6} is not empty and provided a recursive algorithm for constructing the corresponding mollifiers for all $q \in \mathbb{N}$.  He also showed that the Fourier transformation provides a simple characterization of the mollifiers.  But, in the present paper as in most applications of the Colombeau algebras, the explicit knowledge of the form of the Colombeau mollifiers is not necessary:  It is sufficient to know their defining properties \eqref{col:6}. 

 Similarly, as explained in \cite{GSPON2006B}, it is not necessary to know much about the mathematical details of the Colombeau theory to use it in a context such as the present paper.

\section{Interpretation and multiplication of distributions} 
\label{imd:0}

   Distributions $\gamma \in \mathcal{D}'$ are embedded in a Colombeau algebra as the representative sequences $\gamma_\epsilon \in \mathcal{E}_\Mt$ defined by \eqref{col:5} where $\eta$ is a Colombeau mollifier \eqref{col:6}.  Conversely it is possible to recover any distribution $\gamma$ by means of its definition as a functional, i.e., as the equivalence class
\begin{equation}
\label{imd:1}
        \gamma(T) \DEF \lim_{\epsilon \rightarrow 0} 
                 \int dx~\gamma_\epsilon(x) ~T(x),
                \qquad \forall T(x) \in \mathcal{D},
\end{equation}
where $T$ is any test-function, and  $\gamma_\epsilon$ can be any representative of the class $[\gamma] = [\gamma_\epsilon + \mathcal{N}]$ because negligible elements are zero in the limit $\epsilon \rightarrow 0$.

   Of course, as we work in $\mathcal{G}$ and its elements get algebraically combined with other elements, there can be generalized functions $[g_\epsilon]$ different from the class $[\gamma_\epsilon]$ of an embedded distribution which nevertheless correspond to the {same} distribution $\gamma$.  This leads to the concept of \emph{association}, which is defined as follows,\footnote{In the literature the symbol $\approx$ is generally used for association.  We prefer to use $\asymp$ because association is not some kind of an `approximate' relationship, but rather the precise statement that a generalized function corresponds to a distribution.}
\begin{definition}
\label{defi:2}
Two generalized functions $g$ and $h \in \mathcal{G}$, of respective representatives  $g_\epsilon$ and $h_\epsilon$, are said to be associated, and one write $g \ASS h$, iff
\begin{equation}
\label{imd:2}
                 \lim_{\epsilon \rightarrow 0} 
                 \int dx~\Bigl(g_\epsilon(x)-h_\epsilon(x)\Bigr) T(x) = 0,
                 \qquad \forall T(x) \in \mathcal{D}.
\end{equation}
\end{definition}
Thus, if $g$ is a generalized function and $\gamma$ a distribution, the relation $g \ASS \gamma$ implies that $g$ admits $\gamma$ as `associated distribution,' and $\gamma$ is called the `distributional shadow' (or `distributional projection') of $g$ because the mapping $\gamma_\epsilon \mapsto \gamma$ defined by \eqref{imd:1} is then a projection $\mathcal{D}'_\ASS \rightarrow \mathcal{D}'$, where $\mathcal{D}'_\ASS$ such that $\mathcal{D}' \subset \mathcal{D}'_\ASS \subset \mathcal{G}$ is the vector space of elements of $\mathcal{G}$ that have an associated distribution.
 
   The space of distributions is not a subalgebra of $\mathcal{G}$.  Thus we do not normally expect that the product of two distributions in $\mathcal{G}$ will be associated to a third distribution:  In general their product will be a genuine generalized function.

   For example, in $\mathcal{G}$, the square of Dirac's $\delta$-function, Eq.\,\eqref{col:7}, corresponds to $(\delta^2)_\epsilon(x) = (\delta_\epsilon)^2(x) = \epsilon^{-2}\eta^2(-x/\epsilon)$, which has no associated distribution because
\begin{equation} \label{imd:3}
    \lim_{\epsilon \rightarrow 0} \frac{1}{\epsilon}
                 \int \frac{dx}{\epsilon}
            \eta^2\Bigl(\frac{-x}{\epsilon}\Bigr) T(x)
   = \lim_{\epsilon \rightarrow 0} \frac{T(0)}{\epsilon}
                 \int dy~\eta^2(-y)
   = \infty.
\end{equation}
But, referring to \eqref{col:1}, $(\delta^2)_\epsilon(x)$ is a moderate function with $N=2$.  The square of $\delta(x)$ makes thus perfectly sense in $\mathcal{G}$ as a `generalized function' with representative $(\delta^2)_\epsilon(x)=\eta^2(-x/\epsilon)/\epsilon^2$.  Moreover, its point-value at zero, $\eta^2(0)/\epsilon^2$, can be considered as a `generalized number.'  

   On the other hand, we have in $\mathcal{G}$ elements like the $n$-th power of Heaviside's function, Eq.\,\eqref{col:7}, which has an associated distribution but is such that $[\HU^n](x)\neq[\HU](x)$ in $\mathcal{G}$, whereas $\HU^n(x) = \HU(x)$ as a distribution in $\mathcal{D}'$.  Similarly, we have $[x]\odot[\delta](x)\neq 0$ in $\mathcal{G}$, whereas $x\delta(x)=0$ in $\mathcal{D}'$.  In both cases, however, there is no contradiction:  Using \eqref{imd:2} one easily verifies that indeed $[\HU^n](x) \ASS [\HU](x)$ and $[x]\odot[\delta](x)\ASS 0$. 

  These differences between products in $\mathcal{G}$ and in $\mathcal{D}'$ stem from the fact that the distributions embedded and multiplied in $\mathcal{G}$ carry along with them infinitesimal information on the `microscopic structure' of the singularities.  That information is necessary in order that the products and their derivatives are well defined in $\mathcal{G}$, and is lost when the factors are identified with their distributional projection in $\mathcal{D}'$.

  This essential aspect of the product of distributions will be clearly illustrated in the following sections.  It will be seen that whereas the Coulomb field has the simple form $\vec{E}_C = e\vec{r}/r^3$ in the classical theory, its formulation in $\mathcal{G}$ has two terms, i.e., $\vec{E}_\mathcal{G} = e\UPS(r)\vec{r}/r^3 - e\UPS'(r)\vec{r}/r^2$, where $\UPS \ASS \HU$ and $\UPS' \ASS \DUP$ in $\mathbb{R}$.  Therefore, although there is no difference between $\vec{E}_C$ and $\vec{E}_\mathcal{G}$ in any \emph{linear} calculations because $\vec{E}_C \ASS \vec{E}_\mathcal{G}$ in $\mathbb{R}^3$, the usual calculations involving products of $\vec{E}_C$ give inconsistent results, whereas the same calculations with $\vec{E}_\mathcal{G}$ multiplied in $\mathcal{G}$ give mathematically and physically consistent results because of the additional $\delta$-function term in $\vec{E}_\mathcal{G}$.

\section{Point singularities in $\mathcal{D}'$ and in $\mathcal{G}$}
\label{poi:0}

   Classical electrodynamics is basically a continuum theory.  Nevertheless it is possible to consistently introduce point-singularities such as point-charges through distribution theory, and to deal with them successfully, at least as long as the electromagnetic fields, currents, and charge densities interpreted as distributions are not multiplied.  Moreover, it is also possible to deal in specific cases with problems that are non-linear in the fields.  But this requires ad hoc prescriptions, such as defining the energy-momentum tensor, which is quadratic in the fields, directly as a distribution  \cite{TAYLO1956-,ROWE-1978-}.

  In the present paper we are not going to modify anything in Maxwell's theory.  That is, we will use the standard definitions of the potentials, fields, energy-momentum tensor, etc., as well as the standard distributional definitions of the potentials and fields, whose validity can hardly be disputed since they are central in the derivation of the Green's functions of these potentials and fields \cite[p.\,144]{TEMPL1953-}, \cite[p.\,51]{SCHUC1991-}.  Then, since distributions are a subspace of $\mathcal{G}$, we will simply embedded these standard distributions in $\mathcal{G}$ by means of a Colombeau regularization.

   The basic idea in distribution theory is to replace the classical Coulomb potential $e/r$ of a point-charge by the weak limit of the sequence of distributions \cite[p.\,144]{TEMPL1953-},
\begin{equation} \label{poi:1}
  \phi(r) \DEF \lima \frac{e}{r} \HU_a(r),
\qquad \text{where} \qquad
             \HU_a(r) \DEF  \HU(r-a),
\end{equation}
where $e$ is the electric charge of an electron at rest at the origin of a polar coordinate system, and $r = |\vec{r}\,|$ the modulus of the radius vector.\footnote{In references \cite{STEIN1997-} and \cite[p.\,51]{SCHUC1991-} this approach is taken to define the Coulomb \emph{field} rather than the Coulomb potential as the fundamental distribution.  This leads to fields that are equivalent to ours in distribution theory, but which differ by infinitesimals when considered in $\mathcal{G}$.}   Consistent with Schwartz's local structure theorem, $\phi(r)$ is the derivative of $e\lim_{a \rightarrow 0} \log(r/a) \HU(r-a)$, a $\mathcal{C}^0$ function $\forall r \geq 0$.  The infinitesimal cut-off $a > 0$ insures that $\phi(r)$ is a well defined piecewise continuous function for all $r \geq 0$, whereas the classical Coulomb potential $e/r$ is defined only for $r > 0$.  It is then readily verified, referring to \eqref{imd:1}, that \eqref{poi:1} is a distribution.  Indeed,
\begin{equation} \label{poi:2}
   \forall T \in \mathcal{D}, \qquad 
   \iiint_{\mathbb{R}^3} d^3r~ \phi(r) T(r) =
  4\pi e \int_0^\infty dr~ r T(r) \quad \in \mathbb{R},
\end{equation}
because $T \in \mathcal{D}$ has compact support so that the integral is bounded.  

   Differentiating \eqref{poi:2} in the sense of distributions one can calculate the distributional field $\vec{E} = -\vec{\nabla} \phi$ and charge density $\varrho = \vec{\nabla} \cdot \vec{E}$.  But in order to show how this is done in $\mathcal{G}$, and how the resulting $\mathcal{G}$-functions relate to the corresponding distributions, we begin by representing $\phi$ as a mollified sequence according to \eqref{col:5}. The Coulomb potential is thus embedded as the representative sequence
\begin{equation} \label{poi:3}
  \phi_\epsilon(r) = \lima \bigl(\frac{e}{r} \HU_a\bigr)_\epsilon(r)
             = e \int_{\frac{a-r}{\epsilon}}^\infty dy~
                 \frac{\eta(y)}{r+\epsilon y},
                 \qquad \forall r \geq 0.
\end{equation}
One then easily verifies that the distributional Coulomb potential \eqref{poi:1} is recovered when $\epsilon \rightarrow 0$.  Indeed, in that limit, \eqref{poi:3} tends towards $0$ for $r < a$, and towards $e/r$ for $r > a$.  Thus
\begin{equation} \label{poi:4}
  \phi_\epsilon(r) \ASS \lima \frac{e}{r} \HU_a(r) = \phi(r),
\end{equation}
which reverts to the classical Coulomb potential $e/r$ as $a \rightarrow 0$.

  Calculating the embedded Coulomb field is now straightforward because the embedded potential $(e\HU_a/r)_\epsilon$ is $\mathcal{C}^\infty$ in the variable $r$.  It comes
\begin{equation} \label{poi:5}
  \vec{E}_\epsilon(\vec{r}\,) = -\vec{\nabla} \phi_\epsilon(r)
              = e \lima \Bigl(
            \int_{\frac{a-r}{\epsilon}}^\infty dy~
                            \frac{\eta(y)}{(r+\epsilon y)^2}
          - \frac{1}{\epsilon a}\eta\bigl(\frac{a -r}{\epsilon}\bigr)
                        \Bigr) \vec{u},
\end{equation}
where $\vec{u} = \vec{\nabla} r$ is the unit vector in the direction of $\vec{r}$.  Introducing the notation
\begin{equation} \label{poi:6}
   \delta_a(r) \DEF \delta(r-a),
   \qquad \text{so that} \qquad
   \frac{1}{\epsilon a}\eta\bigl(\frac{a -r}{\epsilon}\bigr)
 = \bigl(\frac{1}{a} \DUP_a\bigr)_\epsilon(r),
\end{equation}
this electric field can be written in the more convenient form
\begin{equation} \label{poi:7}
  \vec{E}_\epsilon(\vec{r}\,) = e \lima \Bigl(
            \bigl(\frac{1}{r^2} \HU_a\bigr)_\epsilon(r)
          - \bigl(\frac{1}{a} \DUP_a\bigr)_\epsilon(r)
                                        \Bigr) \vec{u}.
\end{equation}
By an appeal to test functions we easily verify that the field $\vec{E}_\epsilon$ is a distribution, and that the $\DUP$-function in \eqref{poi:7} gives a nul contribution when evaluated on a test function.  Thus
\begin{equation} \label{poi:8}
  \vec{E}_\epsilon(\vec{r}\,) \ASS \lima \frac{e}{r^2} \HU_a(r) \vec{u}
                          \DEF \vec{E}(\vec{r}\,),
\end{equation}
where $\vec{E}(\vec{r}\,)$ is the distributional Coulomb field which in the limit $a \rightarrow 0$ yields the classical Coulomb field $e\vec{r}/r^3$.  Therefore, the distribution $\vec{E}(\vec{r}\,)$ associated to the $\mathcal{G}$-function $\vec{E}_\epsilon$, i.e., its `shadow' obtained by projecting it on $\mathcal{D}'$, does not contain the $\DUP$-function contribution on the right of \eqref{poi:7}.

  To get the Coulomb charge density we have to calculate the divergence of \eqref{poi:5}.  In standard distribution theory one would then ignore the term on the right because it corresponds to a $\DUP$-function which, as we have just seen, gives no contribution when evaluated on a test function.  However, in $\mathcal{G}$, this term cannot be ignored if we subsequently calculate quantities in which $\vec{E}_\epsilon$, or any of its derivatives, is a factor in a product.  The calculation of $\varrho_\epsilon$ is therefore somewhat laborious, but still elementary. It yields, using $\vec{\nabla} \cdot \vec{u} = 2/r$,
\begin{align}
\nonumber
 4\pi \varrho_\epsilon(r) &= \vec{\nabla} \cdot \vec{E}_\epsilon(\vec{r}\,)
              = e \lim_{a \rightarrow 0} \Bigl( ~
         \frac{2}{r} \int_{\frac{a-r}{\epsilon}}^\infty dy~
                            \frac{\eta(y)}{(r+\epsilon y)^2}
        - \frac{2}{\epsilon a r}\eta\bigl(\frac{a -r}{\epsilon}\bigr)\\
 \label{poi:9}
        &+ \frac{1}{\epsilon a^2}\eta\bigl(\frac{a -r}{\epsilon}\bigr)
        -2 \int_{\frac{a-r}{\epsilon}}^\infty dy~
                            \frac{\eta(y)}{(r+\epsilon y)^3}
        + \frac{1}{\epsilon^2 a}\eta'\bigl(\frac{a -r}{\epsilon}\bigr)
                       ~ \Bigr),
\end{align}
which can be rewritten in the less cumbersome form
\begin{align}
\nonumber
 4\pi \varrho_\epsilon(r) &= e \lima \Bigl( ~
        \frac{2}{r} \bigl(\frac{1}{r^2} \HU_a\bigr)_\epsilon(r)
                 -2 \bigl(\frac{1}{r^3} \HU_a\bigr)_\epsilon(r)\\
 \label{poi:10}
        &+ \bigl(\frac{1}{a^2} \DUP_a\bigr)_\epsilon(r)
         - \frac{2}{r}\bigl(\frac{1}{a} \DUP_a\bigr)_\epsilon(r)
         - \bigl(\frac{1}{a}  \DUP'_a\bigr)_\epsilon(r)
                       ~ \Bigr).
\end{align}
The distributional shadow associated to this expression is obtained by evaluating it on a test function \cite{GSPON2006B}, i.e., 
\begin{equation} \label{poi:11}
  \varrho_\epsilon(r) \ASS \lima \frac{e}{4\pi r^2} \DUP_a(r) \DEF \varrho(r),
\end{equation}
which yields the classical point-charge density $ e\delta^3(r) = e \delta(r)/4\pi r^2$ as $a \rightarrow 0$.

\section{$\UPS$-formalism}
\label{ups:0}

Working with regularized distributions can be rather laborious.  However, considerable simplification is possible if the properties of the Colombeau mollifiers \eqref{col:6} are fully taken into account.  In fact, as will be seen in this paper, all typical calculations related to singularities in classical electrodynamics can be made with the help of a generalized function $\UPS(r)$ and its derivatives $\UPS', \UPS'',$ etc., that is
\begin{align} 
\label{ups:1}
   \UPS(r) &\DEF \lim_{a \rightarrow 0} \lim_{\epsilon \rightarrow 0} 
   \bigl(\HU_{a}\bigr)_\epsilon(r) 
 = \lim_{a \rightarrow 0} \lim_{\epsilon \rightarrow 0} 
   \int_{\frac{a-r}{\epsilon}}^\infty dz~ \eta(z),\\
\label{ups:2}
   \UPS'(r) &\,= \lim_{a \rightarrow 0} \lim_{\epsilon \rightarrow 0} 
   \bigl(\DUP_{a}\bigr)_\epsilon(r)
 = \lim_{a \rightarrow 0} \lim_{\epsilon \rightarrow 0} 
   \frac{1}{\epsilon} \eta\Bigl(\frac{a-r}{\epsilon}\Bigr),\\
\label{ups:3}
   \UPS''(r) &\,= \lim_{a \rightarrow 0} \lim_{\epsilon \rightarrow 0} 
   \bigl(\DUP_{a}'\bigr)_\epsilon(r)
 = \lim_{a \rightarrow 0} \lim_{\epsilon \rightarrow 0} 
   \frac{-1}{\epsilon^2} \eta'\Bigl(\frac{a-r}{\epsilon}\Bigr), ...
\end{align}
Indeed, $\UPS$ has properties similar to Heaviside's step-function, and $\UPS'$ to those of Dirac's $\delta$-function, with the fundamental difference that they are $\mathcal{G}$-functions which can be freely multiplied, differentiated, integrated, and combined with any $\mathcal{C}^\infty$ function.  Moreover, the combination $r^{-n}\UPS(r)$ can be given a precise meaning, and the number $0$ in $\mathcal{G}$, i.e., $\OOO (\epsilon^q), \forall q \in \mathbb{N}$, assigned to its point-value at $r=0$.

   Before deriving these properties we recall that both $a$ and $\epsilon$ are in $]0,1[$, that the limit $\epsilon \rightarrow 0$ is to be taken before $a \rightarrow 0$ so that in general $\epsilon/a < 1$, and that the operators $\tfrac{d}{dr}$ and $\int dr$ act before these limits are taken.  For clarity, the variables $x,y,$ and $z$ will be in $\mathbb{R}$, while the variable $r$ will be $\geq 0$.  We also recall that $\eta \in \mathcal{S}$ implies that
\begin{equation} \label{ups:4}
  \lim_{x \rightarrow \pm \infty} |\eta(x)| < \OOO\bigl(|x|^{-q}\bigr),
                                   \qquad \forall q \in \mathbb{N}.
\end{equation}

  We begin by considering $\bigl(\HU_{a}\bigr)_\epsilon(r)$ at fixed $a$ and $r$.  For $r < a$, in which case $(a-r)/\epsilon \rightarrow +\infty$, Eqs.\,\eqref{ups:1} and \eqref{ups:4} yield
\begin{align} 
\nonumber
    \lime \bigl(\HU_{a}\bigr)_\epsilon(r < a) 
 &= \lime 
    \int_{\frac{a-r}{\epsilon}}^{+\infty} dz~ \eta(z)
 <  \lime 
    \int_{\frac{a-r}{\epsilon}\rightarrow + \infty}^{+\infty} dz~ |z|^{-q}\\
\label{ups:5}
 & < \lime  \frac{1}{q+1}\Bigl(\frac{\epsilon}{a-r}\Bigr)^{(q-1)}
   < \OOO\bigl( (\epsilon/a)^q \bigr), \quad \forall q \in \mathbb{N}.
\end{align}
This means that $\bigl(\HU_{a}\bigr)_\epsilon(r < a)$ is negligible, i.e., that $\bigl(\HU_{a}\bigr)_\epsilon(r < a) = 0$ in $\mathcal{G}$, and that in particular $\UPS(0)=0$.   On the other hand, when $r > a$,  the lower bound in \eqref{ups:1} is $(a-r)/\epsilon \rightarrow -\infty$.  Then due to the normalization $\int \eta(z)~dz =1$ we get, using \eqref{ups:5},
\begin{align} 
\label{ups:6}
   \lime \bigl(\HU_{a}\bigr)_\epsilon(r > a) 
 &=          \int_{-\infty}^{+\infty} dz~ \eta(z) 
    - \lime \int_{-\infty}^{\frac{a-r}{\epsilon}\rightarrow -\infty} dz~ \eta(z)
  = 1 - \OOO\bigl( (\epsilon/a)^q \bigr),
\end{align}
so that $\bigl(\HU_{a}\bigr)_\epsilon(r > a) = 1$ in $\mathcal{G}$.

Equations (\ref{ups:5}--\ref{ups:6}) can be written in $\mathcal{G}$ as 
\begin{equation} \label{ups:7}
   \lime \bigl(\HU_{a}\bigr)_\epsilon(r) = 
         \begin{cases}
                      0     &   r < a,\\
                      1     &   r > a,
         \end{cases}
\qquad \text{i.e.,} \qquad
  \Upsilon(r) = 
         \begin{cases}
                      0     &   r = 0,\\
                      1     &   r > 0,
         \end{cases}
\end{equation}
after including the limit $\lima$.  Thus $\UPS$ has indeed properties similar to Heaviside's function, with the essential difference that it has the well defined point-value $\UPS(r)\bigr|_{r=0}=0$.

   We now study the properties of $\bigl(r^{-n}\HU_{a}\bigr)_\epsilon(r)$, i.e.,
\begin{equation} \label{ups:8}
  \lime \bigl(\frac{1}{r^n} \HU_{a}\bigr)_\epsilon(r)
             =   \lime \int_{\frac{a-r}{\epsilon}}^\infty dz~
                 \frac{\eta(z)}{(r+\epsilon z)^n},
                 \qquad \forall n \in \mathbb{Z}.
\end{equation}
This is a $\mathcal{C}^\infty$ expression because the pole at $z=-r/\epsilon < (r-a)/\epsilon$ when $n>0$ is not in the integration range.  We can therefore apply Taylor's theorem, which for $r > a$ enables to write
\begin{equation} \label{ups:9}
  \lime \bigl(\frac{1}{r^n} \HU_{a}\bigr)_\epsilon(r>a)
             =   \frac{1}{r^n} \lime \int_{\frac{a-r}{\epsilon}}^\infty dz~
                 \eta(z)\Bigl(1 - n\epsilon\frac{z}{r}
                + \frac{n(n+1)}{2}\epsilon^2\frac{z^2}{r^2} + ...\Bigr).
\end{equation}
Then, since $(a-r)/\epsilon \rightarrow -\infty$, we can use the Colombeau constraints \eqref{col:6} implying that all terms in the integral with $n=1,...,q$, $\forall q \in \mathbb{N}$, are zero.  Therefore,
\begin{equation} \label{ups:10}
  \lime \bigl(\frac{1}{r^n} \HU_{a}\bigr)_\epsilon(r>a)
             =   \frac{1}{r^n} + \OOO\bigl(\epsilon^q\bigr),
                 \quad \forall q \in \mathbb{N},
\end{equation}
even though the integration in \eqref{ups:9} is $\int^{+\infty}_{\rightarrow -\infty}$ rather than $\int^{+\infty}_{ -\infty}$, because the contribution from this difference is $\OOO (\epsilon^q)$ just like in (\ref{ups:5}--\ref{ups:6}).  Next, in the case $r < a$, for which $(a-r)/\epsilon \rightarrow +\infty$, we
first make the change of variable $y = r+\epsilon z$
\begin{equation} \label{ups:11}
  \lime \bigl(\frac{1}{r^n} \HU_{a}\bigr)_\epsilon(r<a)
             =   \lime \int_{\frac{a-r}{\epsilon}}^\infty 
                 \frac{dz~\eta(z)}{(r+\epsilon z)^n}
             =   \lime \int_a^\infty \frac{dy}{\epsilon}~
                 \eta(\frac{y-r}{\epsilon})\frac{1}{y^n},
\end{equation}
and use \eqref{ups:4} to obtain
\begin{equation} \label{ups:12}
  \lime \bigl(\frac{1}{r^n} \HU_{a}\bigr)_\epsilon(r<a)
             <   \lime \int_a^\infty \frac{dy}{\epsilon}~
                 \frac{\epsilon^q}{(y-r)^q}\frac{1}{y^n}
             <  \OOO\bigl(\epsilon^{q-1}\bigr), \quad \forall q \in \mathbb{N},
\end{equation}
which is equivalent to zero in $\mathcal{G}$, so that in particular $\lime \bigl(r^{-n} \HU_{a}\bigr)_\epsilon(0)=0$.

    In $\mathcal{G}$, Eqs.\,\eqref{ups:10} and \eqref{ups:12} can be merged into the single equation
\begin{equation} \label{ups:13}
   \limea \bigl(\frac{1}{r^n} \HU_{a}\bigr)_\epsilon(r) =
          \begin{cases}
                                                        0   &   r = 0,\\
         r^{-n} \limea \bigl( \HU_{a}\bigr)_\epsilon(r)     &   r > 0,
           \end{cases}
\end{equation}
where (\ref{ups:5}--\ref{ups:6}) enabled to replace $a$ by $0$ on the right, as well as to include $\lima$.  In terms of the $\mathcal{G}$-function $\UPS$, this can be written
\begin{equation} \label{ups:14}
    \bigl(\frac{1}{r^n}  \UPS \bigr)(r) 
        = \frac{1}{r^n}  \UPS(r)
        = \begin{cases}
                      0          &   r = 0,\\
                      r^{-n}     &   r > 0,
           \end{cases}
\end{equation}
provided we agree that the symbol $r^{-n}\UPS(r)$ is a whole and that the assignment $r=0$ is made as on the left-hand side of \eqref{ups:13} so that the point-value $r^{-n}\UPS(r)\bigr|_{r=0}=0$ is well defined.

  Equation \eqref{ups:14} is one of the main practical results of this section: It enables for instance to rewrite the potential \eqref{poi:3} as $e\UPS(r)/r$, and allows to freely associate the powers of $r$ in products such as  
\begin{equation} \label{ups:15}
 r^p \Bigl( \frac{1}{r^n}\UPS(r) \Bigr) \Bigl( \frac{1}{r^m}\UPS(r) \Bigr)
   = \frac{r^p}{r^{n+m}}\UPS^2(r).
\end{equation}

  To obtain similar relations for the derivatives of $\UPS$ we differentiate both sides of \eqref{ups:13}.  For $r=0$ we get of course $0=0$ in $\mathcal{G}$, but for $r>0$ we obtain, after simplification,
\begin{equation} \label{ups:16}
     \limea \bigl(\frac{1}{r^n} \DUP_{a}\bigr)_\epsilon(r)
          = \frac{1}{r^n} \limea \bigl( \DUP_{a}\bigr)_\epsilon(r),
      \qquad \text{or} \qquad 
    \bigl(\frac{1}{r^n} \UPS' \bigr)(r) 
        = \frac{1}{r^n} \UPS'(r),
\end{equation}
as well as, after further differentiations and simplifications,
\begin{equation} \label{ups:17}
    \bigl(\frac{1}{r^n} \UPS^{(m)} \bigr)(r) 
        = \frac{1}{r^n} \UPS^{(m)}(r),
\end{equation}
which is similar to \eqref{ups:14} and shows that $\UPS^{(m)}$ can be multiplied and associated with powers of $r$, just like $\UPS$ in \eqref{ups:15}.

   The calculation of the embedding of $r^{-n}\delta_a$ on the left of \eqref{ups:16} easily leads to
\begin{equation} \label{ups:18}
     \limea \frac{1}{a^n}\bigl( \DUP_{a}\bigr)_\epsilon(r)
          = \frac{1}{r^n} \limea \bigl( \DUP_{a}\bigr)_\epsilon(r),
\end{equation}
which upon differentiation gives
\begin{equation} \label{ups:19}
            \limea \frac{1}{a^n} \bigl( \DUP_{a}'\bigr)_\epsilon(r)
          = \frac{1}{r^n} \limea \bigl( \DUP_{a}'\bigr)_\epsilon(r)
          - \frac{n}{r^{n+1}} \limea \bigl( \DUP_{a}\bigr)_\epsilon(r).
\end{equation}
Equations \eqref{ups:13}, \eqref{ups:18}, and \eqref{ups:19} are useful to reduce expressions in which the parameter $a$ is explicit.  For instance, the charge-density \eqref{poi:10} simplifies to 
\begin{align} \label{ups:20}
  \lim_{\epsilon \rightarrow 0} \varrho_\epsilon(r) 
       = - \limea \frac{1}{r} \bigl(\DUP'_{a}\bigr)_\epsilon(r)
       = - \frac{1}{r} \UPS''(r).
\end{align}
But otherwise Eqs.\,(\ref{ups:18}--\ref{ups:19}) do not bring any particular simplification because when working with the symbols $\UPS, \UPS'$, etc., the infinitesimal $a$ remains implicit.

  The next step is to derive the basic integrations formulas for $\UPS$ and $\UPS'$.  We start with
\begin{align} \label{ups:21}
  \int_0^\infty dr~ \UPS^m(r) ~ F(r)
 = \limea  \int_0^\infty dr~ \bigl(\HU_{a}\bigr)_\epsilon^m(r) ~ F(r),
           \qquad m\in \mathbb{N},
\end{align}
where $F(r)$ stands for
\begin{align} \label{ups:22}
  F(r) = \frac{T(r)}{r^k},
  \qquad \text{with} \qquad
  T \in \mathcal{D}^\infty(\mathbb{R}), \quad k \in \mathbb{N}_0.
\end{align}
Referring to \eqref{ups:7} and \eqref{ups:14}, which specify that apart from the point $r=0$ the function $\UPS(r)$ can be identified with one, the integral on the left of \eqref{ups:21} immediately reduces to
\begin{align} \label{ups:23}
  \int_0^\infty dr~ \UPS^m(r) ~ F(r)
 = \lim_{a \rightarrow 0} \int_a^\infty dr~  F(r),
\end{align}
which is the result anticipated from the similarities between $\UPS$ and the Heaviside function.  As a verification we split the integral on the right of \eqref{ups:21} at $r=a$, and use \eqref{ups:13} to absorb $r^{-k}$ in one of the $\bigl(\HU_{a}\bigr)_\epsilon$ factors
\begin{align} \label{ups:24}
  \limea  \int_0^a dr~
          \bigl(\frac{1}{r^k}\HU_{a}\bigr)_\epsilon(r) \,
          \bigl(\HU_{a}\bigr)_\epsilon^{m-1}(r) \, T(r)
+ \limea  \int_a^\infty dr~
          \bigl(\HU_{a}\bigr)_\epsilon^m(r) \, F(r).
\end{align}
Then, with the help of \eqref{ups:12}, \eqref{ups:5}, and \eqref{ups:6} we get 
\begin{align} \label{ups:25}
  \limea  \int_0^a dr~
          \Bigl(\OOO\bigl( (\epsilon/a)^q \bigr)\Bigr)^m T(r)
+ \limea  \int_a^\infty dr~
          \Bigl(1 - \OOO\bigl( (\epsilon/a)^q \bigr)\Bigr)^m F(r),
\end{align}
which since the $\epsilon$ limit is taken before the $a$ limit confirms \eqref{ups:23}.

  For the $\UPS'$ function we need to evaluate
\begin{align} \label{ups:26}
  \int_0^\infty dr~ (\UPS')^m(r) ~ F(r)
 = \limea  \int_0^\infty dr~ \bigl(\DUP_{a}\bigr)_\epsilon^m(r) ~ F(r),
           \qquad m\in \mathbb{N},
\end{align}
that is, from \eqref{ups:2} and making the change of variable $r = a+\epsilon z$,
\begin{align} \label{ups:27}
  \limea \int_0^\infty dr
   \frac{1}{\epsilon^m} \eta^m\Bigl(\frac{a-r}{\epsilon}\Bigr) F(r)
 = \limea \int_{-a/\epsilon}^\infty dz
   \frac{1}{\epsilon^{m-1}} \eta^m(-z)  F(a+\epsilon z).
\end{align}
Taking  $\epsilon \rightarrow 0$ first the lower limit tends towards $-\infty$.  Then, since $F \in \mathcal{C}^\infty$ for $z > -a/\epsilon$, we use Taylor's theorem and develop the right-hand side as
\begin{align} \label{ups:28}
   \limea \frac{1}{\epsilon^{m-1}}
   \int_{\rightarrow -\infty}^{+\infty} dz~ \eta^m(-z)
   \Bigl( F(a) + \epsilon zF'(a) + \frac{1}{2}\epsilon^2z^2F''(a) + ... \Bigr).
\end{align}
Defining the moments
\begin{align}
 \label{ups:29}
  M[^m_n] = \frac{1}{n!}\int_{-\infty}^{+\infty} dz~ z^n \eta^m(-z),
 \qquad \forall n \in \mathbb{N}_0, \forall m \in \mathbb{N},
\end{align}
we get therefore
\begin{align} 
\nonumber
  \int_0^\infty dr~ (\UPS')^m(r) ~ F(r)
 = \limea \Bigl( M[^m_0]\frac{F(a)}{\epsilon^{m-1}}
               + M[^m_1]\frac{F'(a)}{\epsilon^{m-2}}
             & + M[^m_2]\frac{F''(a)}{\epsilon^{m-3}}\\
\label{ups:30}
               + ...
               + M[^m_{m-1}]  F^{(m-1)}(a)
             & + \OOO(\epsilon) \Bigr),
\end{align}
where for $k\neq 0$ the derivatives $F^{(n)}$ have still to be developed as
\begin{align}
 \label{ups:31}
  \lim_{a \rightarrow 0} F^{(n)}(a) = 
  \lim_{a \rightarrow 0} \Bigl(\frac{T^{(n  )}(0)}{a^k}
                             + \frac{T^{(n+1)}(0)}{a^{k-1}}
                             + ...
                             + \frac{T^{(n+k)}(0)}{k!}
                             + \OOO(a)  \Bigr).
\end{align}
When $k\neq 0$, Eq.\,\eqref{ups:30} leads therefore to a complicated expression with up to $m\times k$ diverging terms as $\limea$.  Nevertheless, for $\UPS'$ we recover the well-known $\delta$-function sifting formula
\begin{align} 
 \label{ups:32}
  \int_0^\infty dr~ \UPS'(r) ~ F(r)
 = \lim_{a \rightarrow 0} F(a),
\end{align}
because $M[^1_0] = 1$ and $M[^1_n] = 0, \forall n \in \mathbb{N}$, due to \eqref{col:6}.  Moreover, for $(\UPS')^2$ we obtain
\begin{align} 
 \label{ups:33}
  \int_0^\infty dr~ (\UPS')^2(r) ~ F(r)
 = \limea  M[^2_0]\frac{F(a)}{\epsilon} + \OOO(\epsilon),
\end{align}
because $M[^2_n] = 0, \forall n$ odd, when $\eta(-z)=\eta(z)$ as in the application considered in the present paper.  Finally, as a consistency check, we integrate by parts the left-hand side of \eqref{ups:32}, and use \eqref{ups:14} and \eqref{ups:23} to confirm that
\begin{align} 
 \label{ups:34}
  \int_0^\infty dr~ \UPS'(r) ~ F(r)
 = \UPS(r) ~ F(r) \Bigr|_0^\infty - \int_0^\infty dr~ \UPS(r) ~ F'(r)
 = \lim_{a \rightarrow 0} F(a).
\end{align}

\noindent\emph{Remark on terminology and notation:}  Because they are associated to the $\delta$-distribution and its derivatives, we will often refer to $\mathcal{G}$-functions like $r^{-n} \UPS', r^{-m} \UPS''$, etc., as `$\delta$-like functions,' `$\delta$-expressions,' or `$\delta$-terms.'  Similarly, we will call $\mathcal{G}$-functions like $r^{-n} (\UPS')^2$, etc., `$\delta^2$-like functions,' etc.  However, in formulas, we will avoid replacing $\UPS'$ by $\delta$ because this may lead to ambiguous expressions unless one works strictly in the linear context.

\section{Electric monopole: Fields and self-energy}
\label{mon:0}

   Let us recalculate the Coulomb field and charge density in the $\UPS$-formalism, and thus replace the definition \eqref{poi:1} of the potential by the equivalent expression \cite{GSPON2004D}
\begin{equation} \label{mon:1}
  \phi(r) \DEF \frac{e}{r} \UPS(r).
\end{equation}
As it is a $\mathcal{C}^\infty$ expression, the electric field can be calculated as in elementary vector analysis.  Thus, since $\vec{\nabla} r = \vec{u}$,
\begin{equation} \label{mon:2}
  \vec{E}(\vec{r}\,)  = -\vec{\nabla} \phi(r)
                      = e \Bigl(
                \frac{1}{r^2} \UPS(r)
              - \frac{1}{r}   \UPS'(r)
                          \Bigr) \vec{u},
\end{equation}
which is fully equivalent to \eqref{poi:7}.   The calculation of $\varrho(r)$ is also elementary, and leads to a greatly simplified result.  Indeed, as $\vec{\nabla} \cdot \vec{u} = 2/r$,
\begin{align}
\nonumber
 4\pi \varrho(r)  &= \vec{\nabla} \cdot \vec{E}(\vec{r}\,)
     = e \Bigl( \frac{1}{r^2}\UPS(r) - \frac{1}{r}\UPS'(r) \Bigr)\frac{2}{r}\\
\label{mon:3}
     &+ e \Bigl(-\frac{2}{r^3}\UPS(r)  + \frac{1}{r^2}\UPS'(r)
               + \frac{1}{r^2}\UPS'(r) - \frac{1}{r}  \UPS''(r) \Bigr)
      = - e \frac{1}{r} \UPS''(r),
\end{align}
which is much simpler than \eqref{poi:10}, and is easily seen to be associated to the usual three-dimensional charge-density because $\UPS''(r)/r \ASS -\delta(r)/4\pi r^2$ in $\mathbb{R}^3$.  Moreover, this expression has the virtue of clearly showing the `origin' of the charge density: The $\UPS(r)$ factor in the potential \eqref{mon:1}. 

   Now that we have formulated the Coulomb potential, field, and charge distributions in $\mathcal{G}$, it is of interest to show that the `extra' $\delta$-like term in $\vec{E}$ given by \eqref{mon:2} or \eqref{poi:7} --- which disappears when these fields are considered as distributions in $\mathcal{D}'$ rather than in $\mathcal{G}$ --- is physically significant.  To do this we calculate the self-energy of a point-charge using the electric field $\vec{E}$ defined according to three theories:  The classical theory, distribution theory, and the $\mathcal{G}$ theory, but using in all three cases the same expression for the self-energy, i.e., the one derived from the standard energy-momentum tensor of the electromagnetic field.  That is, in the classical theory where $\vec{E}(\vec{r}\,)$ is just the Coulomb field $e\vec{r}/r^3$, the integral
\begin{equation} \label{mon:4}
 U_{\text{ele}} \DEF \frac{1}{8\pi} \iiint_{\mathbb{R}^3} d^3r~ \vec{E}^2 
                    = \frac{1}{2} \int_0^\infty dr~ r^2\frac{e^2}{r^4}
                    = \frac{e^2}{2}\lim_{r \rightarrow 0} \frac{1}{r} = \infty.
\end{equation}

    In distribution theory we take for the Coulomb field the distribution $\vec{E}(\vec{r}\,) = \lima e\HU_a\vec{r}/r^3$ defined by \eqref{poi:8}.  Then, apart from expressing the self-energy $U_{\text{ele}}(T)$ as a function of the cut-off $a$, we still have the same divergent result
\begin{equation} \label{mon:5}
 U_{\text{ele}}(1) = \frac{1}{8\pi} \bigl\langle \vec{E}^2 \bigl| 1 \bigr\rangle
                 = \frac{e^2}{2}\lim_{a \rightarrow 0} \frac{1}{a} = \infty,
\end{equation}
even if $\vec{E}^2$ is evaluated on a test-function $T\neq 1$.  Thus, whereas the distribution \eqref{poi:8} is meaningful for all $r \geq 0$, and gives sensible results for expressions linear in $\vec{E}$ evaluated on any test-function, it does not give a sensible result for the self-energy, which is quadratic in  $\vec{E}$. In particular, it is not possible to take the limit $a \rightarrow 0$ which is mandatory for having a \emph{point} charge.

    We now calculate the self-energy in $\mathcal{G}$, where the square of $\vec{E}$ is well defined.  With the Coulomb field expressed as \eqref{mon:2} we have therefore to integrate
\begin{equation} \label{mon:6}
     U_{\text{ele}} 
       = \frac{1}{2} \int_0^\infty dr~ r^2 E^2,
         \qquad \text{where} \qquad
     E \DEF |\vec{E}\,| = \frac{e}{r^2} \UPS(r) - \frac{e}{r} \UPS'(r).
\end{equation}
However, anticipating that similar integrals will have to be integrated in the sequel, we consider the more general integral
\begin{equation} \label{mon:7}
     \mathcal{M}_n = \int_0^\infty dr~ r^n E^2 
         = e^2 \int_0^\infty dr
           \Bigl( r^{n-4}\UPS^2 - 2r^{n-3}\UPS\UPS' + r^{n-2}(\UPS')^2 \Bigr),
\end{equation}
that is, the $n$-th moment of the self-energy-density times $8\pi$. To integrate by parts the $\UPS\UPS'$ term we note that
\begin{equation} \label{mon:8}
\Bigl( r^{n-3}\UPS^2 \Bigr)' =  (n-3)r^{n-4}\UPS^2 + 2r^{n-3}\UPS\UPS',
\end{equation}
so that
\begin{align} \label{mon:9}
     \mathcal{M}_n =  e^2 r^{n-3}\UPS^2\Bigr|_0^\infty
         + e^2 \int_0^\infty dr
               \Bigl( (n-2)r^{n-4}\UPS^2 + r^{n-2}(\UPS')^2 \Bigr),
\end{align}
where the boundary term is zero because of \eqref{ups:14}. Then, using \eqref{ups:23} and \eqref{ups:33} with $F(r)=r^{n-2}$ we get
\begin{align} \label{mon:10}
      \mathcal{M}_n = e^2 \limea \Bigl( \frac{n-2}{3-n}a^{n-3}
                        + \frac{1}{\epsilon}a^{n-2} M[^2_0] \Bigr),
\end{align}
Thus, with $n=2$, we get for \eqref{mon:6}
\begin{equation} \label{mon:11}
 U_{\text{ele}} = \frac{e^2}{2}
                   \lim_{\epsilon \rightarrow 0} \frac{1}{\epsilon} M[^2_0],
\end{equation}
which has the remarkable property to be independent of $a$ as a consequence of the exact cancellation of the  $\UPS^2$ and $\UPS\UPS'$ terms in \eqref{mon:7} for $n=2$.  Eq.\,\eqref{mon:11} is thus rigorously valid in the limit $a \rightarrow 0$, that is for a \emph{point} charge.

  Of course, the self-energy \eqref{mon:11} is infinite in the limit $\epsilon \rightarrow 0$, but originating from a $(\UPS')^2$-term this energy is `concentrated' at the location of the point-charge rather than spread over the field surrounding it as in \eqref{mon:5}.  Moreover, in the $\mathcal{G}$ context, \eqref{mon:11} is a moderate generalized function, more precisely a `generalized number' since it does not depend on any independent variable.  This means that $U_{\text{ele}}$ is mathematically meaningful as a limiting sequence in which $\epsilon$ is kept finite even though $\epsilon \rightarrow 0$.   Consequently, $U_{\text{ele}}$ is a formally diverging quantity which has the attributes required to be subject to `renormalization.'

   Summarizing, when the self-energy is calculated in $\mathcal{G}$ rather than in $\mathcal{D}'$, the $\UPS$ and $\UPS'$ terms in $\vec{E}$ interfere in such a way that in the integral \eqref{mon:6} of $E^2$ the $\UPS\UPS'$-term cancels the $\UPS^2$-term, i.e., the divergent  classical Coulomb-field self-energy \eqref{mon:5}.  The sole contribution to the self-energy comes then from the $(\UPS')^2$ term, which yields the result \eqref{mon:11} that depends only on $\epsilon$ and on the Colombeau mollifier through $M[^2_0] = \int \eta^2(-x)\,dx$, and which may be renormalized to a finite quantity such as the mass of the point-charge. 

   We have therefore obtained the physically remarkable result that calculated in the Colom\-beau algebra the self-energy of a point-charge is entirely located at the position of the charge, and solely due to the square of the $\UPS'(r)$ term in the electric field \eqref{mon:2}, which itself derives form the $\UPS(r)$ factor in the potential \eqref{mon:1}.

\noindent\emph{Remark:} In standard electrodynamics the electrostatic energy of a continuous charge density can be calculated from \cite[p.\,46]{JACKS1975-}
\begin{equation} \label{mon:12}
 U_{\text{ele}} = \frac{1}{2} \iiint d^3x \iiint d^3y 
                   \frac{\varrho(\vec{x}\,) \varrho(\vec{y}\,)}
                        {|\vec{x}-\vec{y}\,|}.
\end{equation}
When applied to a $\delta$-like charge distribution this formula leads to the classical expression \eqref{mon:4}, not to the integral of a $\delta^2$-function as in \eqref{mon:11}.  This highlights a major difference between classical electrodynamics and $\mathcal{G}$-electrodynamics, because in classical electrodynamics the self-energy can be conceived as the energy of merging or separating two `half-charges' at one point, whereas in $\mathcal{G}$-electrodynamics the self-energy of a point-charge is that of a non divisible singularity.  This difference is also related to the classical electrodynamics's concept that the charge density is the source of the electric field at any scale, so that `electron models' are possible, whereas in $\mathcal{G}$-electrodynamics $\varrho$ is not a physical entity on its own but just an attribute of a singularity of the electric field.

\section{Magnetic dipole: Fields and self-energy}
\label{dip:0}

Applying the principles of Sec.~\ref{poi:0} we define the vector potential of a point magnetic dipole as a distribution, and embed it in $\mathcal{G}$.  Therefore
\begin{equation}\label{dip:1}
   \vec A_\epsilon(\vec r\,) \DEF \lima
     \bigl( \frac{\vec{\mu} \times \vec{r}}{r^3} \HU_a \bigr)_\epsilon(r),
   \qquad \text{so that} \qquad
   \vec A(\vec r\,) \DEF
              \frac{\vec{\mu} \times \vec{r}}{r^3} \UPS(r),
\end{equation}
where the magnetic moment $\vec{\mu}$ has the dimension of a charge times a length.  The calculation of the magnetic field strength gives \cite{GSPON2004D}
\begin{equation}\label{dip:2}
    \vec H(\vec r\,) =  \vec\nabla \times \vec A
      = \Bigl( 3\frac{\vec  r}{r^5}(\vec\mu\cdot\vec r)
              - \frac{\vec\mu}{r^3} \Bigr)  \UPS(r)
 +        \frac{\vec{r} \times (\vec\mu\times\vec{r})}{r^4}\UPS'(r),        \end{equation}
which, when $\UPS$ is replaced by $1$ and $\UPS'$ by $0$, reduces to the classical expression
\begin{equation}\label{dip:3}
    \vec H(\vec r\,) = 
        \frac{ 3\vec{r}(\vec{\mu}\cdot\vec{r}) - r^2\vec{\mu} }{r^5}.
\end{equation}
The source current density, given by the rotational of \eqref{dip:2}, is then
\begin{equation} \label{dip:4}
   \frac{4\pi}{c} \vec j(\vec r\,)
    = \vec\nabla \times \vec H(\vec r\,)
    = \Bigl( \frac{2}{r^4} \UPS' - \frac{1}{r^3}\UPS'' \Bigr)
      \vec{\mu} \times \vec{r}
    \ASS 3 \frac{\vec{\mu} \times \vec{r}}{r^4} \DUP(r), 
\end{equation}
which has the usual current density as associated distribution.

   Thus, as in the electric field \eqref{mon:2}, there is a $\delta$-function term in the magnetic field \eqref{dip:2}.  In fact, when integrated over 3-space, this $\DUP$-term gives the well known finite contribution \cite[p.\,184]{JACKS1975-}
\begin{equation}\label{dip:5}
       \iiint d^3r 
       \frac{\vec{r} \times (\vec\mu\times\vec{r})}{r^4}\UPS'(r)
     = \frac{8\pi}{3} \vec\mu ,
\end{equation}
which is essential in calculating the hyperfine splitting of atomic $s$-states \cite{JACKS1977-}. Therefore, contrary to the $\mathcal{G}$-expression \eqref{mon:2} of the electric field of a point-charge, in which the $\delta$-term is not directly observable, we have in the magnetic field of a point dipole a directly observable $\delta$-term \cite{GSPON2004D}. 

   We now calculate the magnetic self-energy of the point-dipole.  Classically, with $\vec{H}$ expressed as \eqref{dip:3}, it is the well-known diverging expression
\begin{equation} \label{dip:6}
 U_{\text{mag}} \DEF \frac{1}{8\pi} \iiint_{\mathbb{R}^3} d^3r~ \vec{H}^2 
                = \frac{\mu^2}{3}\lim_{r \rightarrow 0} \frac{1}{r^3} = \infty.
\end{equation}
In $\mathcal{G}$ we rewrite \eqref{dip:2} as
\begin{equation}\label{dip:7}
  \vec H(\vec r\,) = \vec{V}_1(\vec u\,) \frac{1}{r^3} \UPS(r)
                   - \vec{V}_2(\vec u\,) \frac{1}{r^2} \UPS'(r),
\end{equation}
where, with $\vec{u} = \vec{u} (\theta,\varphi)$ the unit radial vector,
\begin{equation}\label{dip:8}
 \vec{V}_1(\vec u\,) = 3\vec{u}(\vec{\mu}\cdot\vec{u}\,) - \vec{\mu},
     \qquad \text{and} \qquad
 \vec{V}_2(\vec u\,) =  \vec{u}(\vec{\mu}\cdot\vec{u}\,) - \vec{\mu}.     
\end{equation}
Thus
\begin{align}\label{dip:9}
    \vec H^2      = \vec{V}_1^2              \frac{1}{r^6} \UPS^2    
                  - \vec{V}_1 \cdot\vec{V}_2 \frac{2}{r^5} \UPS\UPS'
                  + \vec{V}_2^2              \frac{1}{r^4} (\UPS')^2,
\end{align}
where
\begin{equation}\label{dip:10}
      \vec{V}_1^2 =  3(\vec{u}\cdot\vec{\mu})^2 + \mu^2,
     \qquad \text{and} \qquad
 \vec{V}_1\cdot\vec{V}_2 = \vec{V}_2^2 = \mu^2 - (\vec{u}\cdot\vec{\mu})^2.     
\end{equation}
Then, making the angular integrations with
\begin{align}
\label{dip:11}
    \iint d\omega ~(\vec{u}\cdot\vec{\mu})^2 = \frac{4\pi}{3} \mu^2,
\end{align}
we get
\begin{align}
\label{dip:12}
   U_{\text{mag}} = \mu^2 \int_0^\infty dr
      \Bigl( \frac{1}{ r^4} \UPS^2
           - \frac{2}{3r^3} \UPS\UPS'
           + \frac{1}{3r^2}(\UPS')^2
      \Bigr),
\end{align}
where the first term leads to the classical result \eqref{dip:6} if $\UPS$ is set to one.  To integrate by part we use
\begin{align}
\label{dip:13}
      \Bigl( \frac{1}{3r^3}\UPS^2 \Bigr)'
      = - \frac{1}{r^4}\UPS^2 + \frac{2}{3r^3}\UPS\UPS',
\end{align}
and we obtain
\begin{align}
\label{dip:14}
   U_{\text{mag}} = - \frac{\mu^2}{3r^3}\UPS^2 \Bigr|_0^\infty
                 + \frac{\mu^2}{3} \int_0^\infty  \frac{dr}{r^2}(\UPS')^2,
\end{align}
where the boundary term is zero because of \eqref{ups:14}. Therefore, as in the electric self-energy \eqref{mon:6}, the contributions from the $\UPS^2$ and $\UPS\UPS'$ terms in \eqref{dip:12} cancel each other, and the sole contribution to the magnetic self-energy comes from the $(\UPS')^2$ term, i.e., using \eqref{ups:33} to integrate,
\begin{align}
\label{dip:15}
  U_{\text{mag}} = 
     \frac{1}{3} \mu^2 \lim_{a \rightarrow 0} \lim_{\epsilon \rightarrow 0}
     \frac{1}{a^2} \frac{1}{\epsilon} M[^2_0].
\end{align}
\noindent\emph{Remark:} In standard electrodynamics the magnetostatic energy of a continuous current density can be calculated from a formula \cite[p.\,261]{JACKS1975-} similar to \eqref{mon:12}.  When applied to a $\delta$-like current distribution this formula gives the classical expression \eqref{dip:6}, not the integral of a $\delta^2$-function as in \eqref{dip:15}.

\section{Electron singularity: Self-forces and stability}
\label{ele:0}

Starting in this section we calculate the basic dynamical properties of a classical electron at rest defined as a pole-dipole singularity of the Maxwell field, that is as an electric-pole of charge $e$ and a magnetic-dipole of moment $\vec{\mu}$ located at one same point, taken as the origin of the coordinate system.  To do this we assume that these properties derive solely from the electromagnetic fields generated by that singularity, so that, for instance, the total self-energy of the electron is
\begin{align}
\label{ele:1}
  U_{\text{electron}} = \frac{1}{8\pi} \iiint_{\mathbb{R}^3} d^3r~
      (\vec{E}^2+\vec{H}^2) = U_{\text{ele}} + U_{\text{mag}},
\end{align}
where $U_{\text{ele}}$ and $U_{\text{mag}}$ are the generalized numbers \eqref{mon:11} and \eqref{dip:15}.

  One of the main reasons for considering the electron to be a true point-singularity is that all extended electron models have met with great difficulties, especially with regards to compatibility with special relativity.  A particularly acute problem is that of stability, which arises independently of relativistic transformations in static models already, because the repulsive character of all electrical forces implies that an extended electron modeled as a small sphere or shell will necessarily tend to fly apart.  

    A general solution to this problem consists of introducing `Poincar\'e compensating stresses' of a non-electromagnetic nature \cite{POINC1906-,BIALY1979-}.  Such stabilizing forces are also required in point-like electron models in which the energy-momentum tensor is interpreted as a Schwartz distribution \cite{ROWE-1978-,LOZAD1989-}.  Moreover, Poincar\'e stresses can be given an interpretation in the context of general relativity \cite{BIALY1979-}, which may provide a final answer to the questions of self-stress and self-energy of charged particles in arbitrary motion \cite{TANGH1962-}.

   In the present paper we only deal with the question of stability in the rest-frame of the singularity.  What matters therefore is the total self-force, i.e.,
\begin{align}
\label{ele:2}
  \vec{F}_{\text{electron}} =  \iiint_{\mathbb{R}^3} d^3r~
      (\varrho\vec{E} + \vec{j} \times \vec{H}) = 0,
\end{align}
where $\vec{E}$ and $\vec{H}$ are given by \eqref{mon:2} and \eqref{dip:2}, and  $\varrho$ and $\vec{j}$ by \eqref{mon:3} and \eqref{dip:4}.  It then turns out that this force is zero.  Indeed, when calculating in $\mathcal{G}$ all integrations are made with $\epsilon \neq 0$ and $a \neq 0$, so that all quantities being finite the radial and angular integrations can be interchanged without their order affecting the result.  Then, as the electric force density $\bigl(\varrho\vec{E}\bigr)(r\vec{u}\,)$ changes sign in the substitution $\vec{u} \rightarrow - \vec{u}$, the electric self-force is zero upon angular integration; similarly, as the magnetic force density $\bigl(\vec{j} \times \vec{H}\bigr)(r\vec{u}\,)$ also changes sign in that substitution, the integrated magnetic force is also zero.  Since the electron is strictly point-like, this implies that the forces which potentially tend to its disassembly compensate each other exactly at one \emph{single} point:  The electron is therefore stable and Poincar\'e compensating stresses are not required.

  To confirm this conclusion with an explicit computation, and to unveil some interesting features of the radial self-force, we calculate the electric self-force, i.e.,
\begin{equation} \label{ele:3}
 \vec{F}_{\text{ele}} \DEF \iiint_{\mathbb{R}^3} d^3r~
                            \varrho(\vec{r}\,)\vec{E}(\vec{r}\,) 
                    =  \iint d\omega ~\vec{u}(\theta,\varphi) F_{\text{r}},
\end{equation}
where writing $E(r) = |\vec{E}(\vec{r}\,)|$ the radial self-force is given by
\begin{equation} \label{ele:4}
 F_{\text{r}} \DEF \frac{d|\vec{F}_{\text{ele}}|}{d\omega}
               = \int_0^\infty dr~ r^2 \varrho(r) E(r),
\end{equation}
In principle $F_{\text{r}}$ could be calculated using Eqs.\,(\ref{mon:2}--\ref{mon:3}) for $\vec{E}$ and $\varrho$.  But a more convenient method is to remark, referring to \eqref{mon:3}, that one can write $\varrho= 2E/r + E'$ so that the integral becomes
\begin{equation} \label{ele:5}
 F_{\text{r}} =  \int_0^\infty dr~ r^2 E \Bigl(\frac{2}{r} E + E'\Bigr).
\end{equation}
This enables to integrate by parts using the identity
\begin{equation} \label{ele:6}
      E \Bigl(\frac{2}{r} E + E'\Bigr)
   = \frac{1}{2r^4} \Bigl( 4r^3 E^2 + 2r^4 E E' \Bigr)
   = \frac{1}{2r^4} \Bigl( \bigl(r^2 E^2\bigr)^2 \Bigr)',
\end{equation}
which is well defined at $r=0$ because of \eqref{ups:14}. Thus
\begin{equation} \label{ele:7}
 F_{\text{r}} =  \frac{1}{2} r^2E^2 \Bigr|_0^\infty
              + \int_0^\infty dr~ r E ^2,
\end{equation}
where the boundary term is zero on account of \eqref{ups:14}.  Therefore,
\begin{equation} \label{ele:8}
    F_{\text{r}}  =  \int_0^\infty dr~ r E^2 = \mathcal{M}_1,
\end{equation}  
i.e., from \eqref{mon:10},
\begin{equation} \label{ele:9}
 F_{\text{r}}   = e^2 \lim_{a \rightarrow 0} \lim_{\epsilon \rightarrow 0}
                  \Bigl( \frac{1}{a\epsilon} M[^2_0] - \frac{1}{2a^2} \Bigr)
                > 0.
\end{equation}
Consistent with \eqref{ele:7} this force is positive, i.e., such that $\vec{F}_{\text{ele}}$ is outwards directed, because  $\epsilon \ll a$ since the $\epsilon$ limit is taken first.  Thus, as $\epsilon \rightarrow 0$, the $\delta^2$-divergence in the self-force prevails, and the negative term (which comes from the interference between the $\UPS$ and $\UPS'$ terms in $\vec{E}$ and $\varrho$) has only an infinitesimally small stabilizing effect.  Similar conclusions can be drawn for the radial self-force due to the magnetic dipolar field. 

   Consequently, the electric and magnetic self-forces are effectively concentrated as $\delta^2$-divergences at the location of the electron singularity.  Then, despite that these forces are outwards directed, the symmetries of the electric and magnetic fields imply that the angular integrations lead to a zero total self-force.  This confirms the stability of the punctual electron.

\section{Self-momentum and self-angular-momentum}
\label{sam:0}

The total electromagnetic-field momentum and angular-momentum are defined as
\begin{align}
\label{sam:1}
 \vec{P} &\DEF \frac{1}{4\pi c} \iiint_{\mathbb{R}^3} d^3r~
              \vec{E} \times \vec{H},\\
\label{sam:2}
 \vec{S} &\DEF \frac{1}{4\pi c} \iiint_{\mathbb{R}^3} d^3r~
              \vec{r} \times (\vec{E} \times \vec{H}\,).
\end{align}
When the fields are due to purely static sources the momentum $\vec{P}$ is called `hidden momentum,' and when these sources correspond to a singularity such as a single electron, $\vec{P}$ is called self-momentum, and $\vec{S}$ self-angular-momentum or `spin.'  

   For example, a system consisting of a point magnetic-dipole of moment $\vec{m}$ located at $\vec{r} = 0$, and a point charge $q$ positioned at $\vec{r} = \vec{s}$, has a non-zero hidden momentum \cite{GSPON2007C, FURRY1969-, AHARO1988-}
\begin{eqnarray}\label{sam:3}
    \vec{P}
      =  \dfrac{q}{s^3} \vec{m} \times \vec{s},
\end{eqnarray}
which tends towards infinity as $1/s^2$ when the point charge approaches the position of the point dipole.  We therefore expect that the hidden momentum of an electron singularity, i.e., a point-charge $e$ located at the same position as a point-dipole $\vec{\mu}$, could also be infinite --- unless it is zero as in the case of the self-force.  On the other hand, due to the $\vec{r}$ factor in \eqref{sam:2}, the self-angular-momentum of an electron singularity could be less divergent.  To find out we have to calculate $\vec{P}$ and $\vec{S}$.

   Taking for $\vec{E}$ and $\vec{H}$ expressions \eqref{mon:2} and (\ref{dip:7}--\ref{dip:8}), we get, since $\vec{u}\times\vec{u}=0$,
\begin{align}\label{sam:4}
     \vec{E} \times \vec{H} = \frac{\vec{\mu} \times \vec{u}}{er} E^2,
\end{align}
where $E(r)$ is the modulus of $\vec E$ as in \eqref{mon:6}.  The momenta \eqref{sam:1} and \eqref{sam:2} are then
\begin{align}
\label{sam:5}
 \vec{P} &= \frac{1}{4\pi ec} \iint d\omega~ 
                \vec{\mu} \times \vec{u} 
               ~\int_0^\infty dr~rE^2,\\
\label{sam:6}
 \vec{S} &= \frac{1}{4\pi ec} \iint d\omega~
                \vec{u} \times ( \vec{\mu} \times \vec{u}\,)
               ~\int_0^\infty dr~r^2E^2,
\end{align}
in which the radial integrals are moments of the form \eqref{mon:10}.  As for the angular integrals, since the angular part of \eqref{sam:5} is odd in the substitution $\vec{u} \rightarrow -\vec{u}$, its integral is zero.  On the other hand, as the angular part of \eqref{sam:6} is even, its integral is non-zero, i.e., $8\pi\vec{\mu}/3$.   Separating the angular and radial integrals and writing $P_r=\mathcal{M}_1$ and $S_r=\mathcal{M}_2$, we have therefore
\begin{align}
\label{sam:7}
               \vec{P} &= \vec{I}_\omega P_r, \qquad \text{where} \qquad
                          \vec{I}_\omega = 0,\\
\label{sam:8}
               \vec{S} &= \vec{J}_\omega S_r, \qquad \text{where} \qquad
                          \vec{J}_\omega = \frac{2}{3ec} \vec{\mu},
\end{align}
and, using \eqref{mon:10},
\begin{align}
\label{sam:9}
     P_r &= e^2 \lim_{a \rightarrow 0} 
                \lim_{\epsilon \rightarrow 0} \Bigl(
                \frac{1}{a\epsilon} M[^2_0]  -\frac{1}{2a^2} \Bigr),\\
\label{sam:10}
     S_r &= e^2 \lim_{\epsilon \rightarrow 0}
                \frac{1}{\epsilon} M[^2_0].
\end{align}

   The radial component of the self-momentum has therefore the same form as the radial component of the self-force \eqref{ele:9}, i.e., a prevailing $\delta^2/a$-divergence with a negligible $1/a^2$ correction.  But, as was the case for the self-force, the angular integral being zero implies that the total self-momentum is zero.

   On the other hand, the self-angular-momentum is independent on $a$ and a pure $\delta^2$-divergence, just like the electric self-energy \eqref{mon:11}, i.e.,
\begin{align}
\label{sam:11}
     \vec{S} = \frac{2e}{3c} \vec{\mu} 
               \lim_{\epsilon \rightarrow 0}
               \frac{1}{\epsilon} M[^2_0],
\end{align}
where the $\delta^2$ comes from the product of the $\delta$-terms in $\vec{E}$ and $\vec{H}$.

\section{Renormalization of mass and spin}
\label{ren:0}

   In the previous sections we have found that the basic dynamical quantities which can be calculated for an electron singularity at rest are either zero, i.e., the self-force \eqref{ele:2} and the self-momentum \eqref{sam:7}, or else $\delta^2$-divergences, i.e., the self-energy \eqref{ele:1} and the self-angular-momentum \eqref{sam:11}.  These non-zero quantities are therefore entirely located at the position of the electron, i.e., precisely where its dynamical attributes as a point-mass are supposed to reside.  This enables to identify them with the mechanical mass and spin of a point-electron, although this is only possible after renormalization since these non-zero quantities are actually infinite in the limits $\epsilon \rightarrow 0$ and $a \rightarrow 0$.  But the requirements $\epsilon \neq 0$ and $a \neq 0$ are also the conditions under which the fields of the electron and the quantities deriving from them make sense as $\mathcal{G}$-functions and $\mathcal{G}$-numbers:  The Colombeau theory of nonlinear generalized functions is therefore congenial to renormalization!

    In order to do this renormalization we proceed as in any field-theory.  That is, we begin by identifying the free parameters, which must be of the same number as the physically measurable quantities for the theory to be renormalizable.  In our case, the diverging quantities are \eqref{mon:11}, \eqref{dip:15}, and \eqref{sam:11}, i.e.,
\begin{align}
\label{ren:1}
  U_{\text{ele}}(\epsilon) &= \frac{1}{2} \frac{M[^2_0]}{\epsilon}
                              e^2,\\
\label{ren:2}
U_{\text{mag}}(a,\epsilon) &= \frac{1}{3} \frac{M[^2_0]}{\epsilon}
                              \frac{\mu^2}{a^2},\\
\label{ren:3}
         \vec{S}(\epsilon) &= \frac{2}{3} \frac{M[^2_0]}{\epsilon}
                              \frac{e\vec{\mu}}{c},
\end{align}
which we have rewritten as functions of the parameters $a$ and $\epsilon$, because $a$ and $\epsilon$ are the two free parameters of our theory since the quotient $M[^2_0]/\epsilon$ can be considered as a simple rescaling of $\epsilon$.  This enables to renormalize two quantities, which we obviously take as the mass and the spin, and which can be written in terms of $U_{\text{ele}}(\epsilon)$ as 
\begin{align}
\label{ren:4}
      \bigl(mc^2\bigr)(a,\epsilon) &=  U_{\text{ele}} + U_{\text{mag}}
  = \Bigl(1 +\frac{2}{3}\frac{\mu^2}{e^2a^2} \Bigr) U_{\text{ele}}(\epsilon),\\
\label{ren:5}
        S(\epsilon) &= \frac{4}{3} \frac{\mu}{ec}
                               U_{\text{ele}}(\epsilon).
\end{align}
Then, taking for $m$ and $S = |\vec{S}\,|$ any positive number, we can solve these equations for $a$ and $\epsilon$, also assumed to be positive, which will be acceptable solutions provided they are such that $\epsilon \ll a$.

  This is all that can be done within the exclusive context of dynamics:  To go further additional input is required from classical electrodynamics and quantum theory.

\section{Consistency with quantum theory}
\label{cqt:0}

   In classical electrodynamics the magnetic-moment $\mu$ and the angular-momentum $S$ of a closed system of moving charges and circulating currents are related by the equation \cite[p.\,183]{JACKS1975-}
\begin{align}
\label{cqt:1}
    \mu = \frac{e}{2mc} S.
\end{align}
In the quantum theory of elementary particles, this equation is replaced by
\begin{align}
\label{cqt:2}
    \mu = \frac{e\hbar}{2mc} g s,
\end{align}
where $s= 0, \tfrac{1}{2}, 1, \tfrac{3}{2}, ...$ is the quantum mechanical spin in units of $\hbar$, and $g$ is the gyromagnetic ratio which according to Telegdi's conjecture equals $g=2$ for truly elementary particles of any spin \cite{TELEG1992-}.  Since these equations contain the mass $mc^2=U_{\text{ele}} + U_{\text{mag}}$, whereas \eqref{ren:5} contains only $U_{\text{ele}}$,
they lead to a relation between $U_{\text{ele}}$ and $U_{\text{ele}} + U_{\text{mag}}$ which constrains the admissible values of $g$.  Indeed, combining \eqref{ren:5} and \eqref{cqt:2}, we get
\begin{align}
\label{cqt:3}
    mc^2 = \frac{2}{3}gU_{\text{ele}},
\end{align}
which is compatible with \eqref{ren:4} only if $g \geq 3/2$.

  Consequently, the classical relation \eqref{cqt:1}, which corresponds to $g=1$, does not apply to the spin of the electron singularity considered in this paper.  This was of course to be expected since there is no moving charge or circulating current in a static singularity, what confirms that our results are consistent with classical electrodynamics.

  It remains therefore to investigate whether our results are also consistent with  quantum theory.  Thus we use \eqref{cqt:2} with $g=2$ to rewrite \eqref{ren:5} as
\begin{align}
\label{cqt:4}
   S(a,\epsilon)  = \frac{4}{3} \frac{U_{\text{ele}}(\epsilon)}{mc^2} s\hbar,
\end{align}
so that, using \eqref{ren:4}, we are led to the equation
\begin{align}
\label{cqt:5}
   S(a,\epsilon)  = \frac{4e^2a^2}{3e^2a^2 + 2\mu^2}  s.
\end{align}
This equation is consistent, i.e., $S = s\hbar$, iff 
\begin{align}
\label{cqt:6}
     ea = \pm \frac{1}{\sqrt{2}}\mu,
\end{align}
which by \eqref{cqt:2} also implies that
\begin{align}
\label{cqt:7}
        a = \frac{1}{\sqrt{2}}\frac{\hbar}{mc} s.
\end{align}
Then, putting \eqref{cqt:6} in \eqref{ren:4}, we find  $mc^2 = \tfrac{4}{3} U_{\text{ele}}$ and thus, from \eqref{ren:1},
\begin{align}
\label{cqt:8}
        \epsilon = \frac{2}{3} M[^2_0] \frac{e^2}{mc^2}
                 = \frac{2}{3} M[^2_0] \alpha \frac{\hbar}{mc},
\end{align}
where $\alpha = e^2/\hbar c \approx 1/137$.

   Therefore, the cut-off parameter $a$ is on the order of the Compton wave-length $\hbar/mc$, i.e., on the same order as the `Zitterbewegung' radius characteristic of Dirac's electron theory.  On the other hand, as $M[^2_0] \approx 1$, the regularization parameter $\epsilon$ is on the order of the classical electron radius $e^2/mc^2$, which is $\approx 137$ times smaller than the Compton wave-length, so that $\epsilon \ll a$.

   The results obtained in this paper are thus consistent with quantum theory and the concept that the magnetic moment and spin of an electron are not due to moving charges or circulating currents.  Moreover, the magnitude $a \approx \hbar/mc$ confirms that quantum mechanical corrections to classical electrodynamics become essential at distances below $\hbar/mc$ --- that is, by \eqref{cqt:1}, at angular momenta smaller than $2\hbar$ per unit charge --- and the magnitude $\epsilon \approx \alpha \hbar/mc$ implies that quantum-field-theoretical corrections become essential at distances below $\alpha\hbar/mc$.

\noindent \emph{Remark:} To conclude this section it is perhaps important to stress that the finite values attributed to $a$ and $\epsilon$ in the renormalization process do not mean that $a$ and $\epsilon$ are not \emph{infinitesimal} as has been assumed throughout the paper.  Quite the contrary,  firstly because the Maxwell field is scale invariant so that any given finite value can be considered as infinitesimal, and, more fundamentally, because this apparent contradiction is in the nature of the renormalization process.  Indeed, the mathematical singularities considered in this paper are in agreement with the notion that the electron is truly point-like, as is experimentally confirmed to extreme precision.  Thus, had we studied these singularities in the context of quantum field theory we would have found more complicated relations to be compared to more precise data on a scale much smaller than $a$ and $\epsilon$.  But again we would have been confronted with the issue that below that scale there is `new' physics that we do not yet understand.

\section{Conclusion}
\label{con:0}

The main objective of this paper was to use the formalism of Colombeau generalized functions to properly define the electric-monopolar and magnetic-dipolar fields of a point-electron singularity in the Maxwell field, i.e.,
\begin{equation}
\label{con:1}
  \vec{E}(\vec{r}\,) = \frac{e\vec{r}}{r^3} \UPS (r)
                      - \frac{e\vec{r}}{r^2}   \UPS'(r),
\end{equation}
%
%
\begin{equation}
\label{con:2}
    \vec H(\vec{r}\,)
      = \frac{3\vec{r} (\vec\mu\cdot\vec{r}) - \vec\mu}{r^5}\UPS (r)
      + \frac{ \vec{r} \times (\vec\mu\times\vec{r})  }{r^4}\UPS'(r),        \end{equation}
and to calculate the basic dynamical characteristics of this pole-dipole singularity, what required a framework such that the multiplication of distributions is possible, because these characteristics are quadratic in the fields.

   The main conclusion is that the $\UPS'$-terms in the fields \eqref{con:1} and \eqref{con:2} are mathematically meaningful and physically significant.  In particular, despite that these $\delta$-like-terms can generally be ignored in linear operations on the fields, they are essential to calculate the most fundamental dynamical attributes of an electron considered as a point-mass: Mass and spin.

   Moreover, these $\delta$-terms also enter into the calculation of the self-force and self-momentum, which however turn out to be zero so that the electron singularity is stable.

   While the electric charge $e$ and magnetic moment $\vec{\mu}$ of the electron are finite integrals of $\delta$-functions (i.e., the charge density $\varrho$ and the current density $\vec{j}$), its mass $m$ and spin $\vec{S}$ are integrals of $\delta^2$-functions, i.e., the self-energy density $(\vec{E}^2+\vec{H}^2)/8\pi$ and the self-angular-momentum density $\vec{r} \times (\vec{E} \times \vec{H}\,)/4\pi c$, which have to be renormalized.

   More precisely, the only contributions to the mass coming from the electric field \eqref{con:1} and magnetic field \eqref{con:2}  are from the squares of the $\delta$-terms in these fields, whereas the spin is entirely due to the product of the $\delta$-term in  $\vec{E}$ times the $\delta$-term in $\vec{H}$.

  Mass and spin are therefore entirely located at the position of the electron singularity, in agreement with the classical notion of mechanical inertia of a point-mass. 

   Finally, the fact that mass and spin, i.e., self-energy and self-angular-momen\-tum, are integrals of $\delta^2$-expressions subject to renormalization makes them suitable to interpretation as classical limits in a quantum field theory of electrons.


\section{References}
\label{biblio:0}

\end{document}